\documentclass[preprint,aps,showpacs]{revtex4}
\usepackage{graphicx}

\def\be7pg{$^7Be(p,\gamma)^8B$}
\def\xbe7{$^7Be$}
\def\b8{$^8B$}
\def\S17{$S_{17}(0)$}
\def\xs17{$S_{17}$}
\def\s34{$S_{34}(0)$}
\def\xpm{$\pm$}

\begin{document}

\title{Comment on Reconciling Coulomb 
 Dissociation and Radiative Capture Measurements}
\thanks{Work Supported by USDOE Grant No. DE-FG02-94ER40870.}

\author{Moshe Gai}
\affiliation{Laboratory for Nuclear Science at Avery Point, 
University of Connecticut, 
and Yale University.}


\maketitle

Esbensen, Bertsch and Snover \cite{PRL} 
suggested that higher order Coulomb Interactions and an E2 correction are 
important to Coulomb Dissociation (CD) of \b8 \cite{PRL}. It is claimed that 
"\xs17 values extracted from CD data have a significant steeper slope as a 
function of $E_{rel}$, the relative energy of the proton and the \xbe7 fragment, 
than the direct result". Hence they reanalyze CD data and claim that corrections 
of the original analyses yield slope values in better agreement with Direct 
Capture (DC) data.

Specifically they calculate a very large (20\%) correction for the RIKEN2 
\cite{Kik} data at the lowest energy and suggest a substantial (50\%) correction 
of the b-slope parameter. The corrections of other CD data are small 
and in some case(s) vanish due to fortuitous cancellation \cite{PRL}. 
They  imply for the RIKEN2 data "slope correction similar in 
magnitude to the 0.25 $MeV^{-1}$ average slope difference between CD and direct 
results as shown in Fig. 19 of \cite{Jung03}". Note that here they refer to the b-slope fit 
parameter (in units of $MeV^{-1}$) used in \cite{Jung03}, \xs17(E) = a(1+bE), 
and not the usual physical slope S' = dS/dE. 

In Fig. 1 we show the RIKEN2 \xs17 data \cite{Kik}  
using analysis employing first order Coulomb Dipole (E1) interaction only, 
and compare it for example to the Seattle data \cite{Jung03} on DC. These data 
were also compared to DC data in \cite{Sch03} from which it is clear that the slope of the 
RIKEN2 data is in agreement with DC data. Kikuchi {\em et al.} on the other hand  
\cite{Kik} already emphasized an agreement with the DC data available at that time.
We observe in Fig. 1 good agreement between the slope of the published 
RIKEN2 CD data above 300 keV (S' = 6.4 \xpm 1.5 eV b/MeV and b = 0.4 \xpm 
0.1 $MeV^{-1}$) and the slope of the Seattle DC data (S' = 5.8 \xpm 0.6 eV b/MeV and
b = 0.32 \xpm 0.02 $MeV^{-1}$).

From Fig. 1 it is also clear that the different b-slope plotted in Fig. 19 of  \cite{Jung03} is 
due to their use of a subset of the RIKEN2 data and a neglect of the systematic error (8.6\%)
discussed by Kikuchi {\em et al.} \cite{Kik}. The five data points shown in Fig. 1 
(for $E_{rel}$ = 375, 625, 875, 1125 and 1375 keV, $S_{17}$ = 17.48(171), 
19.84(108), 21.44(105), 22.52(230), 24.13(164), eV-b, respectively)  include the (8.6\%) systematic 
error  discussed in \cite{Kik} or a slightly smaller systematic 
error. In Ref. \cite{Kik} a less refined systematic error   
with only one value (8.6\%) is quoted. No systematic errors were included in the fit of \cite{Jung03}.  As discussed in \cite{Kik} these  systematic errors are due mainly to the subtraction of the background from the dissociation in the helium bag which varies among data points, see Fig. 1 of our Letter publication \cite{Kik}. Such a (varying) systematic error must be considered for each data point separately (in the same way that one considers background subtracted from a peak in a spectrum). 

\begin{figure}
\includegraphics[width=5in]{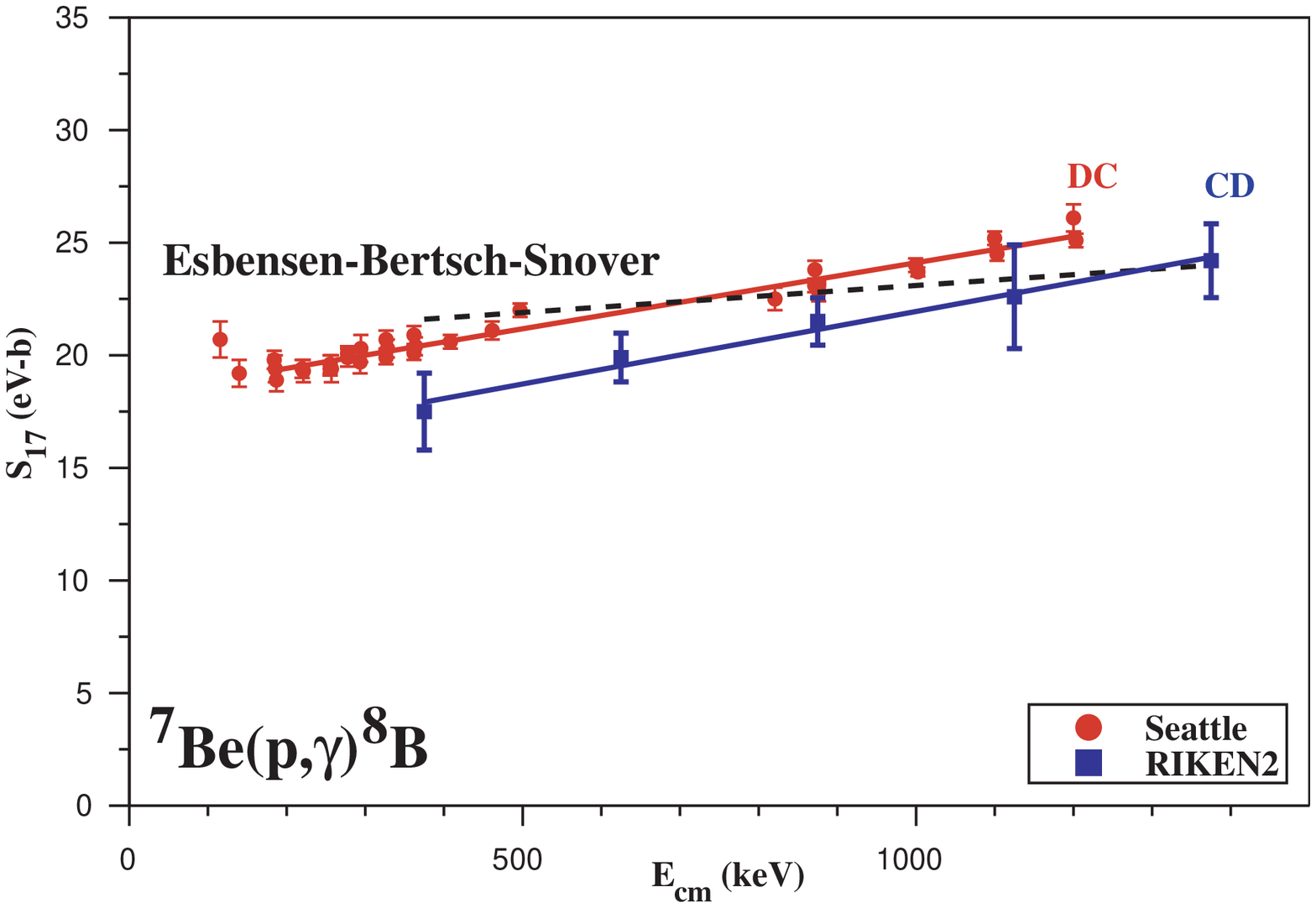}
 \caption{\label{RIKEN2} (Color online) Extracted \xs17 from the RIKEN2 CD data \cite{Kik} using first 
order electric dipole interaction as shown in \cite{Sch03}, compared to the DC 
capture data published by the Seattle group \cite{Jung03}. The shown RIKEN2 data include 
systematic uncertainties (equal or slightly smaller) as published  \cite{Kik}.}

\end{figure} 

For the RIKEN2 \cite{Kik} published data (b = 0.4 \xpm 0.1  $MeV^{-1}$) 
the implied correction of 0.25 $MeV^{-1}$  \cite{PRL}, 
yields b = 0.15 \xpm 0.1 $MeV^{-1}$, more than a factor of 2 smaller 
than the so called average b-slope of DC data \cite{Jung03}.
The corrected RIKEN2 data are not shown but discussed in Ref. \cite{PRL},
where it is stated that \xs17 is increased by 20\% at low 
energy and slightly smaller at $E_{rel} \ \approx \ 1.375$ MeV. In Fig. 1 we 
show the so described slope with a dashed line. The corrected 
slope is smaller than the slope of DC (e.g. Seattle) data. The 
proposed corrections \cite{PRL} in fact lead to a disagreement 
and do not reconcile the slopes of the RIKEN2 CD and DC data.

\end{document}